\documentclass[12pt]{report} 
\vspace{-1.5cm}
\usepackage{epsfig}
\usepackage{citesort}

\begin{document}
\title{A Second Look at Canonical Sampling of Biomolecules using Replica Exchange Simulation}
\author{
Daniel M. Zuckerman and Edward Lyman 
\\
\normalsize
Department of Computational Biology, School of Medicine, \\ 
\normalsize
and Department of Environmental \& Occupational Health, \\
\normalsize
Graduate School of Public Health, \\ 
\normalsize
Suite 3064 BST3, 3501 Fifth Avenue, University of Pittsburgh,
\normalsize
Pittsburgh, Pennsylvania 15213 \\
\normalsize
Phone: 412-648-3335; Fax: 412-648-3163 \\
\normalsize
dmz@ccbb.pitt.edu 
}
\normalsize
\date{(\today)}
\maketitle

Because of growing interest in temperature-based sampling methods like replica exchange \cite{Swendsen-1986,Hansmann-1997,Garcia-2002,Garcia-2004,dePablo-2005b,Simmerling-2005,Duan-2005}, this note aims to make some observations and raise some potentially important questions which we have not seen addressed sufficiently in the literature.  
Mainly, we wish to call attention to limits on the maximum speed-up to be expected from temperature-based methods, and also note the need for careful quantification of sampling efficiency. 
Because potentially lengthy studies may be necessary to address these issues, we felt it would be useful to bring them to the attention of the broader community.  
Here \emph{we are strictly concerned with canonical sampling at a fixed temperature,} and \emph{not} with conformational search.

We will base our discussion around a generic replica exchange protocol, consisting of $M$ levels spanning from the temperature $T_0$ at which canonical sampling is desired, up to $T_M$.  
The protocol is motivated by the increased rate of barrier crossing possible at higher temperatures.  
We assume each level is simulated for a time $t_{\mbox{sim}}$, which implies a total CPU cost $(M+1) \times t_{\mbox{sim}}$.  
In typical explicitly solvated peptide systems, $M \sim 20$, $T_0 \simeq 300K$ and $T_M \sim 450K$ \cite{Garcia-2002} [check temp].  
The relatively low $T_M$ values reflect the well-known, high sensitivity of the approach to configuration-space overlap in large systems \cite{Hansmann-1997,Garcia-2002}:
that is, because of minimal overlap, typical configurations in high $T$ Boltzmann ensembles are unlikely in low $T$ enembles.
We note that a new exchange variant introduced by Berne and coworkers permits the use of ``cold'' solvent and larger temperature gaps \cite{Berne-2005}, but the issues we raise still apply to the new protocol, especially as larger solutes are considered.

While replica exchange is often thought of as an 'enhanced sampling method,' what does that mean?  Indeed, what is an appropriate criterion for judging efficiency?  
As our first observation, we believe \textbf{(Obs.\ I)} efficiency can only mean a decrease in the total CPU usage --- i.e., \emph{summed over all processors} --- for a given degree of sampling quality.  
(We will defer the necessary discussion of assessing sampling quality, and only assume such assessment is possible.)   
When the goal is canonical sampling at $T_0$, after all, one has the option of running an ordinary parallel simulation at $T_0$ (e.g., [namd]) or even $M$ independent simulations \cite{Karplus-1998}.  
A truly efficient method must be a superior alternative to such ``brute force'' simulation.

\textbf{(Obs.\ II)} Reports in the literature offer an ambiguous picture as to whether replica exchange attains efficiency for canonical sampling.  
Sanbonmatsu and Garcia compared replica exchange to an equivalent amount of brute-force sampling, but their claim of efficiency is largely based on the alternative goal of enhancing sampling over the full range of temperatures, rather than for canonical sampling at $T_0$ \cite{Garcia-2002}.  
When the data solely for $T_0$ are examined, there is no clear gain, especially noting that assessment was based on principal components derived only from the replica exchange data.  
Another claim of efficiency, by Duan and coworkers \cite{Duan-2005}, fails to include the full CPU cost of all $M$ levels.  
When suitably corrected, there does appear to be speedup of perhaps a factor of two for $T_0 = 308K$, but the system studied is considerably smaller (permitting larger temperature jumps) than would be possible in protein systems of interest.  
Another efficiency claim by Roe \emph{et al.} also does not account for the full CPU cost of all ladder levels \cite{Simmerling-2005}.
In a structural-glass system, replica exchange was found not to be helpful \cite{Sciortino-2002}, although efficiency has been noted in spin-systems \cite{Swendsen-1986,Swendsen-2005b}.
We emphasize that \emph{biomolecular} replica exchange should indeed be efficient in certain cases (with high enough energy barriers; see below).  
At least one such instance has been noted by Garcia, using a suitable brute-force comparison system \cite{Garcia-2005}.

The lack of clear-cut results in a much-heralded approach merit closer examination.  
What might be preventing efficiency gain?  Or put another way, what is the maximum efficiency possible in a standard replica exchange simulation?  
The very construction of the method implies that \textbf{(Obs.\ III)} in any parallel exchange protocol, the sampling ``speed'' at the bottom level --- lowest $T$ --- will be controlled by the speed at which the top level --- highest $T$ --- samples the necessary space.  
Further, given our interest in efficient canonical sampling at $T_0$, the speed of the top level should exceed that of the bottom by \emph{at least} a factor of $M$.  
If not, the simulation does not ``break even'' in total CPU cost, as compared to brute-force canonical sampling at $T_0$ for the full length $M \times t_{\mbox{sim}}$.

The basic temperature dependence of barrier-crossing is well known (e.g., \cite{Atkins-2002}) and has important consequences for replica exchange.  
The Arrhenius factor indicates that the temperature-dependent rate $k$ for crossing a particular barrier obeys
\begin{equation}
\label{arrhenius}
k_a(T) = k_0 \exp{ \left( + \Delta S / k_B \right) } \exp{ \left( -\Delta E / k_B T \right) } \, 
\end{equation}
for a fixed-volume system, where $k_0$ is an unknown prefactor insensitive to temperature and assumed constant; $\Delta E$ is the energy barrier and $\Delta S$ is the entropy barrier --- i.e., "narrowing" of configuration space --- which must be expected in a multi-dimensional molecular system.  
Two observations are immediate: \textbf{(Obs.\ IV)} the entropic component of the rate is completely unaffected by an increase in temperature, and the possible speedup due to the energetic part can easily be calculated.  

The table gives possible speedups for several energy barriers and temperatures, employing units of $k_B T_0$ for $T_0=300K$.  
Speed-ups are computed simply as the ratio $k_a(T_M) / k_a(T_0)$ for possible values of $T_M$.  
It is clear that for modest barriers, the speed-up attainable even with a top temperature $T_M = 500K$ is only of the order of a typical number of replicas in replica exchange, $M \sim 20$.  
Thus, \textbf{(Obs.\ V)} if modest barriers ($< 8 k_B T_0$) dominate a system's dynamics, efficiency will be difficult to obtain via replica exchange, since the speed-up noted in the table needs to be divided by $M+1$.  

How high are barriers encountered in molecular systems?  
We can only begin to answer this question, but one must first be careful about which barriers matter.  
We believe that \textbf{(Obs.\ VI)} ``local'' barriers will matter most: that is, the energy barriers actually encountered in a trajectory will dominate sampling speed.  
Apparent barriers determined by projections onto arbitrary low-dimensional reaction coordinates would seem of uncertain value.  
(We note that Zwanzig has attempted to account for local roughness with an effective diffusion constant on a slowly varying landscape \cite{Zwanzig-1988}.)

Evidence from simulations and experiments is far from complete, but indicates that \textbf{(Obs.\ VII)} energy barriers in molecular systems appear to be modest.  
Here, unless noted otherwise, $T_0 \simeq 300K$.
In their extensive study of a tetrapeptide, Czerminski and Elber found barriers $<$ 3 kcal/mole $\simeq 5 k_B T_0$ for the lowest energy transition path \cite{Elber-1990}.
Equally interesting, they found approximately 1,000 additional paths with similar energy profiles (differing by $<$ 1 kcal/mole $< 2 k_B T_0$) --- suggesting what we might term a ``pebbly'' rather than ``mountainous'' energy landscape.
See also Ref.\ \cite{Thirumalai-2003}.
In our own work (unpublished) with implicitly solvated dileucine, increasing the temperature from 298K to 500K led to a hopping-rate increase of a factor of 1.8, suggesting a small barrier ($< 1.5 k_B T_0$).  
Similarly, Sanbonmatsu and Garcia found that barriers for explicitly solvated met-enkephalin were small, on the order of $k_B T_0$ \cite{Garcia-2002}.  
An experimental study has also suggested barriers are modest ($< 6 k_B T_0$) \cite{Reich-2005}.  
Although this list is fairly compelling, we believe the question of barrier heights is far from settled.
Further study should carefully consider local vs.\ global barriers, as well as entropy vs.\ energy components of barriers.  
(We purposely do not discuss barriers to protein folding, because our scope here is solely equilibrium fluctuations.)

Finally, the goal of understanding efficiency implies the need for reliable means for assessing sampling.  
An ideal approach to assessment would survey all pertinent substates to ensure appropriate Boltzmann frequencies.
Present approaches to assessment typically calculate free energy surfaces (equivalently, probability distributions) on one or two-dimensional surfaces, which are evaluated visually.  
Principal components (e.g., \cite{Garcia-2002,Duan-2005}) as well as ``composite'' coordinates like the radius of gyration \cite{Simmerling-2005} are popular coordinate choices.  
Yet we believe that \textbf{(Obs.\ VIII)} the use of low-dimensional sampling assessment is intrinsically limited, since it could readily mask structural diversity --- i.e., be consistent with substantially distinct conformational ensembles.
Future work could usefully pursue higher-dimensional measures, which can always be numerically compared between independent simulations for sampling assessment.
In our own work, for instance, we have begun to use a histogram measure which directly reports on the structural distribution of an ensemble \cite{Zuckerman-2006b}.

In conclusion, we have attempted to tie together a number of straightforward observations which reflect concerns about the effectiveness of the replica exchange simulation method, when the goal is single-temperature canonical sampling.  
The concerns suggest other simulation strategies, such as Hamiltonian exchange \cite{Okamoto-2000} and resolution exchange \cite{Zuckerman-2006a,Zuckerman-2006c}, may merit consideration --- as well as scrutiny.  
We emphasize that our goal has been to raise questions more than to answer them.  
Even if our worries turn out to be exaggerated, a candid discussion of the issues should be beneficial to the molecular simulation community.

The authors wish to thank Rob Coalson, Juan de Pablo, Ron Elber, Angel Garcia, and Robert Swendsen for very useful conversations.
We gratefully acknowledge support from the NIH, through Grants ES007318 and GM070987.
We also greatly appreciate support from the Department of Computational Biology and the Department of Environmental \& Occupational Health. 

\newpage
\begin{table}
\nonumber
\begin{center}
\begin{tabular}{r|cccc}
\hline \hline
             & $\Delta E = 2 k_BT_0$ & $4 k_B T_0$ & $6 k_B T_0$ & $8 k_B T_0$\\
\hline
$T_M=400K$   &  1.65                 &  2.72       &  4.48       &  7.39      \\
$500K$       &  2.23                 &  4.95       &  11.0       &  24.5      \\
$600K$       &  2.72                 &  7.39       &  20.1       &  54.6      \\
\hline \hline
\end{tabular}
\end{center}
\caption{
High-temperature speed-up factors calculated using Arrhenius factors.
Speed-up factors are computed as the ratio $k_a(T_M) / k_a(T_0\!=\!300K)$ for the indicated energy barriers $\Delta E$ via Eq.\ (\ref{arrhenius}).
Energy barriers are given in units of $k_B T_0$.
A rough estimate of the efficiency factor (the factor by which total CPU usage is reduced) obtainable in an $M$-level parallel replica exchange simulation with maximum temperature $T_M$ is the table entry divided $M$.
}
\end{table}
\pagebreak

\end{document}